\documentclass[sigconf]{acmart}

  \renewcommand\footnotetextcopyrightpermission[1]{} 
\usepackage{booktabs} 
\usepackage{url}  
\usepackage{graphicx}  
\usepackage{color}
\usepackage{hyperref}
\usepackage{tikz}
\usepackage{pgfplots}
\usepackage{float}
\usepackage{amsmath,amssymb}
\usepackage{color}
\usepackage{booktabs}
\usepackage{multirow}
\usepackage{bbm}
\usepackage[linesnumbered,ruled]{algorithm2e}
\usepackage{caption}
\usepackage{subcaption}

\setcopyright{rightsretained}

\newcommand{\etal}{\textit{et al}.}
\newcommand\T{\rule{0pt}{2.9ex}}       
\newcommand\B{\rule[-1.2ex]{0pt}{0pt}} 

\begin{document}

\title{Causal Embeddings for Recommendation}

\author{Stephen Bonner}
\authornote{Also a doctoral student at the Department of Computer Science, Durham University, UK}
\orcid{1234-5678-9012}
\affiliation{%
  \institution{Criteo AI Labs}
  \city{Paris}
}
\email{st.bonner@criteo.com}

\author{Flavian Vasile}
\affiliation{%
  \institution{Criteo AI Labs}
  \city{Paris}
}
\email{f.vasile@criteo.com}

\renewcommand{\shortauthors}{S. Bonner \& F. Vasile}

\begin{abstract}
Many current applications use recommendations in order to modify the natural user behavior, such as to increase the number of sales or the time spent on a website. This results in a gap between the final recommendation objective and the classical setup where recommendation candidates are evaluated by their coherence with past user behavior, by predicting either the missing entries in the user-item matrix, or the most likely next event. To bridge this gap, we optimize a recommendation policy for the task of increasing the desired outcome versus the organic user behavior. We show this is equivalent to learning to predict recommendation outcomes under a fully random recommendation policy. To this end, we propose a new domain adaptation algorithm that learns from logged data containing outcomes from a biased recommendation policy and predicts recommendation outcomes according to random exposure. We compare our method against state-of-the-art factorization methods, in addition to new approaches of causal recommendation and show significant improvements.
\end{abstract}

%
%

\copyrightyear{2018} 
\acmYear{2018} 
\setcopyright{acmlicensed}
\acmConference[RecSys '18]{Twelfth ACM Conference on Recommender Systems}{October 2--7, 2018}{Vancouver, BC, Canada}
\acmBooktitle{Twelfth ACM Conference on Recommender Systems (RecSys '18), October 2--7, 2018, Vancouver, BC, Canada}
\acmPrice{15.00}
\acmDOI{10.1145/3240323.3240360}
\acmISBN{978-1-4503-5901-6/18/10}
  
  

\maketitle

\section{Introduction}
\label{sec:intro}
In recent years, online commerce has outpaced the growth of traditional business. The research work on recommender systems has consequently grown significantly during the same time-frame. As shown by key players in the online e-commerce space, such as Amazon, YouTube or Netflix, the product recommendation functionality is now a key driver of demand, accounting for, in the case of Amazon, roughly $35\%$ \cite{amzreco} of overall sales.

Within the area of recommender systems, a promising new class of deep-learning based solutions has emerged and are showing promising results. We can separate the newly proposed solutions in two main types of approaches: ones that learn item embeddings that optimize for item-item similarity prediction \cite{prod2vec,metaprod2vec,glove} and the ones that learn user sequence embeddings to optimize for next item prediction \cite{hidasi2015session,hidasi2016parallel}. The new methods can scale to millions of users and items and show good performance improvements over traditional approaches. In the context of product recommendation at scale, these deep-learning approaches were successfully applied to ad recommendations in Yahoo! Mail \cite{prod2vec}, for video recommendation at YouTube \cite{covington2016deep} and restaurant recommendations by OpenTable \cite{opentable}.

However, even with the recent model developments, the current state-of-the-art machine learning models still frame the recommendation task as either:
\begin{itemize}
\item A distance learning problem between pairs of products or between pairs of users and products, where we measure outcomes using link prediction/matrix completion metrics such as Mean Squared Error (MSE) and Area Under the Curve (AUC).

\item A next item prediction problem that models the user behavior and tries to predict what she will do next, using ranking metrics like Precison@K and Normalized Discounted Cumulative Gain (NDCG).
\end{itemize}

Both of the approaches fall short of modeling the intrinsic interventionist nature of recommendation, which should not only attempt to model the organic user behavior, but to actually optimally influence it according to a present objective. In other words, in many cases, recommendations are by design actions that have, as a common desired outcome, the change in user behavior, e.g. an increase in a certain user activity such as buying products, watching movies or applying for a mortgage/credit card.

Ideally, the increase in user activity would be measured against the baseline case where no recommendations are being shown. Using a causal vocabulary, \emph{we are interested in finding the optimal treatment recommendation policy that maximizes the reward with respect to the control recommendation policy for each user, also known as the Individual Treatment Effect (ITE) \cite{rubin1974estimating}}. In this context, the control policy constitutes the current state of the system. This can be either a system with no recommendations in place or a baseline recommender system that we want to improve using the currently logged feedback. In the rest of the paper, we will use control and logging policy interchangeably to describe the baseline system.

Currently, this type of approach is prevalent in fields such as personalized medicine, where one tries to infer which treatment would have the biggest positive impact for a particular patient. Similar questions are starting to be asked in the field of Performance Advertising \cite{johnson2015ghost}, where the problem of measuring the incremental effect of an advertising campaign on the user shopping propensity is very important for proper credit attribution in the context of single and multi-channel advertising \cite{berman2016beyond}. In terms of associated rewards, this marks a natural evolution of the field from measuring advertising performance in terms of number of ad displays (CPM - cost per thousand ad displays), to measuring it in terms of ad clicks (CPC), to post-click sales/conversions (CPA) to possibly incremental sales (incremental CPA - value of a sale times the associated individual treatment effect).

In our paper, we introduce a simple modification of standard matrix factorization methods that leverages a small sample of randomized recommendations outcomes in order to create user and products representations. For our method, the associated pairwise distance between user and item pairs is more aligned with the corresponding ITE than in traditional approaches. In Section \ref{sec:experiments}, we show that, because of this, our method can lead to significant lifts in performance over both classical and recent causal inference-based recommendation methods.

\subsection{Learning Recommendation Policies Optimized for ITE}

In order to introduce our method, we adopt a notation similar to the one introduced in \cite{swaminathan2015batch} and \cite{rosenfeld2016predicting} and extend it to the matrix factorization case. The primary notation used throughout this section is detailed in Table \ref{tab:notation}.

\begin{table}[!h]
  \centering
  \caption{Definitions and Notations}
  \label{tab:notation}
  \begin{tabular}{c  p{6.5cm}}
    \toprule
   \textbf{Symbol} & \textbf{Definition} \\
   \midrule \midrule
   $u_i$ & A user of the recommendation system \T \\
   $p_j$ & A product which the system can recommend \\
   $\pi_x$ & A recommendation policy (eq. \ref{eq:policy}) \\
   $\pi_c$ & The control recommendation policy. This represents the recommendation system used to create the training dataset.\\
   $\pi_t$ & The treatment recommendation policy. This represents the updated recommendation system.\\   
   $\pi^{rand}$ & The fully random recommendation policy that shows any product with equal probability to all users.\\
   $r_{ij}$ & The true reward for recommending a product $p_j$ to user $u_i$ \\
   $y_{ij}$ & The observed reward for recommending product $p_j$ to user $u_i$ in the data. By comparison with $r_{ij}$, its value can be unknown.\\
   $R^{\pi_x}$ & The total reward for policy $\pi_x$ (eq. \ref{eq:policyreward}) \\
   $ITE_{ij}^{\pi_x}$ & The difference between the reward for current and control policy (eq. \ref{eq:itepair}) \\
   $p_i^*$ & The product with the highest reward for user $u_i$ (eq. \ref{eq:optimalprod}) \\
   $\pi^*$ & The best incremental recommendation policy (eq. \ref{eq:optimalpolicy}) \\
   $S_c$ & A large set of training samples collected under the control recommendation policy \\
   $S_t$ & A smaller set of samples taken under a full-randomized recommendation policy \B \\
   \bottomrule
  \end{tabular}
\end{table}

\paragraph{Recommendation Policy}
Consider a recommender system that takes as input a user $u_i \in \mathcal{X}$, from the user population $\mathcal{X}$ and outputs as its prediction one of the possible recommendable products $p_j \in \mathcal{P}$. The external behavior of the recommender system can be described by the policy of which product to show for each user. We assume a stochastic policy $\pi_{x}$ that associates to each user $u_i$ and product $p_j$ a probability for the user $u_i$ to be exposed to the recommendation of product $p_j$:
\begin{equation}
\label{eq:policy}
p_j \sim \pi_{x}(.|u_i)
\end{equation}
For simplicity we assume showing no products is also a valid intervention in $\mathcal{P}$.

\paragraph{Policy Rewards}
We define $r_{ij}$ to be the true outcome/reward for recommending product $p_j$ to user $u_i$. In our case, $r_{ij}$ is a binary outcome, for example click/no click, sale/no sale. We assume that the reward $r_{ij}$ is distributed according to an unknown conditional distribution $r$ depending on $u_i$ and $p_j$:
$$
r_{ij} \sim r(.|u_i,p_j)
$$

We define $y_{ij}$ to be the observed reward for the pair $i,j$ of user-product according to the logging policy $\pi_x$:
$$
y_{ij} = r_{ij}\pi_x(p_j | u_i)
$$ 

The reward $R^{\pi_x}$ associated with a policy $\pi_x$ is equal to the sum of the rewards collected across all incoming users by using the associated personalized product exposure probability:
\begin{equation}
\label{eq:policyreward}
R^{\pi_x} =  \sum_{ij}  r_{ij} \pi_x(p_j | u_i) p(u_i) =  \sum_{ij}  y_{ij} p(u_i)  = \sum_{i} R_{ij} 
\end{equation}
where the probability of seeing users comes from an unknown distribution $p(\mathcal{X})$: $u_i \sim p(\mathcal{X})$ and $R_{ij}$ is the reward associated with the user-product pair $ij$.

The Individual Treatment Effect (ITE) value of a policy for a given user $i$ and a product $j$ for a policy $\pi_x$ is defined as the difference between it's reward and the reward of the control policy $\pi_c$:

\begin{equation}
\label{eq:itepair}
ITE_{ij}^{\pi_x} =  R_{ij}^{\pi_x} - R_{ij}^{\pi_c}
\end{equation}

In our paper, we are interested in finding the policy $\pi^{*}$ with the highest sum of ITEs:
\begin{equation}
\label{eq:optimalpolicy}
\pi^{*} = arg\max_{\pi_x} \{ITE^{\pi_x} \}
\end{equation}
where: $ITE^{\pi_x} = \sum_{ij} ITE_{ij}^{\pi_x}$

\paragraph{Lemma 1}

For any control policy $\pi_c$, the best incremental policy $\pi^*$ is the policy that shows deterministically to each user the product with the highest associated reward.
\paragraph{Proof}
Let $\pi^*$ be the policy with the highest associated reward:
\begin{equation}
\pi^{*} = arg\max_{\pi_x} \sum_{ij} r_{ij} \pi_x(p_j | u_i) p(u_i)
\end{equation}
We have that $\pi^{*}$ is the policy with the highest ITE w.r.t to any control policy $\pi_{c}$ since $ITE^{\pi^*} = R^* - R_c \geq R^{\setminus *} - R_c$.

Let $p_i^*$ be the product with highest reward for user $u_i$:
\begin{equation}
\label{eq:optimalprod}
p_i^* = arg\max_{p_j} r_{ij}
\end{equation}
and $r_i^*$ the associated reward.

For any arbitrary policy $\pi_x$ we have the following inequality that holds:
\begin{equation}
\begin{split}
R^{\pi_x} =  \sum_{i} p(u_i) \sum_{j} r_{ij} \pi_x(p_j | u_i) \leq \\ \leq \sum_{i} p(u_i) \sum_{j} r_i^* \pi_x(p_j | u_i) = \sum_{i} p(u_i) r_i^* = R^{\pi_{det}}
\end{split}
\end{equation}
where $\pi_{det}$ is the deterministic policy of showing the best personalized product to each user $u_i$:
\begin{equation}
\pi_{det}=
\begin{cases}
    1,& \text{if } p_j = p_i^{*} \\
    0,              & \text{otherwise}
\end{cases}
\end{equation}
Therefore, $\pi^* = \pi_{det}$. \hspace{4.7cm} $\square$ \newline

In order to find the optimal policy $\pi^*$, we need to find for each user $u_i$ the product with the highest personalized reward $r_i^{*}$. However, in practice we do not observe directly $r_{ij}$, but $y_{ij} \sim r_{ij} \pi_x(p_j | u_i)$. Most of the current approaches use \emph{Inverse Propensity Scoring (IPS)}-based methods (see \cite{liangcausal} and \cite{schnabel2016recommendations}) to predict the unobserved reward $r_{ij}$:
\begin{equation}
\hat{r}_{ij} \approx \frac{y_{ij}}{\pi_c(p_j | u_i)}
\end{equation}
The main shortcoming of these approaches is that IPS-based estimators do not handle well big shifts in exposure probability between treatment and control policies \cite{bottou2013counterfactual} (products with low probability under the logging policy $\pi_c$ will tend to have higher predicted rewards).

Ideally, for minimum variance, the logging policy $\pi_c$ should be one of uniform exposure recommendations (denoted as $\pi^{rand}$), which is impossible in practice due to the resulting low recommendation quality. One potential solution is then to use the biased data from the current $\pi_c$ and learn to predict the outcomes under a randomized policy and then rank the products by their predicted outcomes under this transformation. More formally, if we denote by $L(.)$ the loss associated with our policy learning objective, we have that:
\begin{equation}
min \sum_{u_i} p(u_i) L(\hat{p_i}, p_i^{*})
\end{equation}
Assuming $\hat{p_i}$ is computed as the maximum over predicted rewards we have that:
\begin{equation}
\begin{split}
 \hat{p_i} = arg\max_{p_j} \hat{r}_{ij} = arg\max_{p_j} \frac{\hat{y}_{ij}}{\pi_c(p_j | u_i)} = \\
 = arg\max_{p_j} \frac{\hat{y}^{rand}_{ij}}{\pi^{rand}(p_j | u_i)} = arg\max_{p_j} \hat{y}_{ij}^{rand}
 \end{split}
\end{equation}
where $\hat{y}^{rand}_{ij}$ is the predicted outcome of showing product $p_j$ to user $u_i$ under randomized exposure and
all $\pi^{rand}(p_j | u_i)$ values are equal by the definition of randomized exposure.

Therefore, if we can build a good predictor for $\hat{y}_{ij}^{rand}$, we can avoid the explosion in variance we encounter in the classic IPS approach.

The key difference between domain adaptation and IPS-based methods is that in order to be able to learn a model that performs well on the fully randomized domain, we need a sample from it. That means that at learning time we assume the existence of two training samples: $S_c = { \{ (u_i, p^c_j, y^c_{ij}) \} }^{M_c}_{i=1}$, that is a very large sample of exposed users with outcomes collected with the control recommendation policy and $S_t = { \{ (u_i, p^t_j, y^t_{ij}) \} }^{M_t}_{i=1}$, that is a much smaller sample of exposed users with outcomes collected with the fully randomized recommendation policy (for example, collected using an e-greedy exploration strategy \cite{sutton1998reinforcement} or from other exploration strategies that can be used to simulate outcomes from $\pi^{rand}$ by rejection sampling). A new class of embedding-based approaches which leverage this setup have been recently proposed in \cite{johansson2016learning,shalit2016estimating} and the closely related work in \cite{rosenfeld2016predicting}.

In our work, we leverage their findings and propose \emph{CausalEmbed (CausE)}, a new embedding method that is able to transfer evidence between a large sample of logged feedback under the control policy $S_c$ and a small sample of logged treatment policy feedback $S_t$ in order to be able to better predict randomized treatment effects on pairs of users and products. We show that our approach significantly improves recommendation performance on unseen pairs of users and products exposed according to the randomized treatment policy $\pi^{rand}_t$.

The key contributions of our work are the following:
\begin{itemize}
\item We extend the past work \cite{rosenfeld2016predicting} on learning from biased control plus randomized treatment and make the connection with the recent work on causal embeddings \cite{shalit2016estimating} and propose a new approach entitled \emph{CausE}, a new matrix factorization algorithm that generalizes both previous approaches.
\item We apply the newly introduced algorithm on the problem of product recommendation and benchmark it against standard factorization and IPS-based causal methods and analyze various ways of leveraging the exploration sample $S_t$ in order to improve the performance of the model on unseen test data from the randomized treatment policy.

\end{itemize}

To aid the reproducibility of our \emph{CausE} method, we release the source code\footnote{\url{https://github.com/criteo-research/CausE}} and present results only on public benchmark datasets. 

The structure of the paper is the following: In Section \ref{sec:relatedwork} we cover previous related work and the relationship with our method. In Section \ref{sec:proposedapproach} we present the \textit{CausE} approach. In Section \ref{sec:experiments} we present the experimental setup and the results on the MovieLens dataset. In Section \ref{sec:conclusion}, we summarize our findings and conclude with future directions of research.


\section{Related Work}
\label{sec:relatedwork}
The body of work focusing upon causality is vast and diverse, spanning multiple interconnected disciplines. We concentrate our review on the work measuring causal effects as a counterfactual inference problem.

\subsection{Propensity Scoring Methods}

The basic idea of propensity scoring methods is to turn the outcomes of an observational study into a pseudo-randomized trial by re-weighting samples, similarly to importance sampling. This is the classical \emph{Inverse Propensity Scoring (IPS)} \cite{rosenblum1983central} method of computing an unbiased estimator of the reward of policy $\pi_t$ under its own induced distribution from a sample of data collected under policy $\pi_c$.

In \cite{bottou2013counterfactual} the authors introduce \emph{Clipped Inverse Propensity Scoring}, a method which addresses the variance problems that the unbiased IPS estimator is facing when $\pi_t$ and $\pi_c$ are too different.

Additionally, \cite{dudik2011doubly} introduces \emph{Double Robust estimation for IPS}, as a way to evaluate the value of a new policy on logged data that combines the direct method estimation with the IPS estimation and provides an estimator which is consistent if at least one of the two models is well-specified.

In a series of papers \cite{swaminathan2015counterfactual,swaminathan2015batch}, the authors propose the \emph{Counterfactual Risk Minimization (CRM)} principle along with a new learning setup \emph{Batch Learning from Bandit Feedback (BLBF)} that aims to answer the counterfactual question of: "How will a system perform in response to an intervention that changes the data distribution?". As a solution, they propose the \emph{Policy Optimizer for Exponential Models (POEM)} algorithm, that optimizes a variance-regularized version of the Clipped-IPS estimator. In the context of learning under the BLBF setup, \cite{lefortier2016large} releases a dataset with logged propensity weights according to a baseline policy $h_0$ and benchmarks various propensity weighting methods using the CRM principle. 

The most recent work in the BLBF setup introduces \emph{BanditNet}, an algorithm for training deep neural networks via logged contextual bandit feedback data \cite{joachims2-18deep}. This approach views the output of a neural network to be analogous with that of a a stochastic policy $\pi_w$, where $w$ are the parameters associated with the model. Using BanditNet, $\pi_w$ can be trained in the following manner:
$$
\hat w = arg\min_w \frac{1}{n} \sum_{i=1}^{n} (\delta_i-\lambda) \frac{\pi_w(p_j|u_i)}{\pi_c(p_j|u_i)}
$$
where $\pi_c$ is an existing training policy, $\delta_i$ is the received loss for $p_j$ given $u_i$ under $\pi_c$ and $\lambda$ is an unconstrained Lagrange multiplier, which must be chosen empirically.
\subsection{Causal Inference using Domain Adaptation and Transfer Learning Methods}
\emph{Causal inference as Domain Adaptation} \cite{johansson2016learning,shalit2016estimating}: The authors frame the problem of causal inference as a domain adaptation problem where we need to learn on the factual domain and predict on the counterfactual domain. By leveraging the results on Domain Adaptation from \cite{courty2016optimal}, the authors show that the expected ITE estimation error of a representation is bounded by the sum of the supervised generalization error of the representation and the distance between the resulting treated and control distributions in the new representation.

\emph{Causal inference as Transfer Learning/Joint Optimization} \cite{rosenfeld2016predicting}: The authors propose a new method that allows for better prediction of the magnitude of a new treatment effect by leveraging not only the A/B Test traffic where the traffic is exposed to the new treatment, but also the pre-A/B Test data that was exposed to the control policy.

\subsection{Causal Recommendations}
The current work in causal recommendations focuses on the idea of using causal inference as a way to de-bias matrix factorization on existing logged data. In \cite{liang2016modeling} the authors present ExpoMF, a probabilistic approach for collaborative filtering on implicit data. ExpoMF jointly models both users exposure to an item, and their resulting click decision, resulting in a model which naturally down-weights the expected, but ultimately un-clicked items. The model is also able to consider additional covariates which may impact the exposure, for example a user's location when recommending restaurants. In \cite{liangcausal} the authors follow-up with a second version of the model for rating matrices and with explicit modeling of the exposure model (they propose two exposure models, one of global popularity and one that personalizes the exposure for each user). Once the exposure model is estimated the preference model is fit with weighted click data, where each click (or skip) is weighted by the inverse probability of exposure (coming from the exposure model). As a result, the observational click data is weighted as though it came from an "experiment" where users are randomly shown items. This leads to good performance on the domain-adaptation task where the model is estimated on a different distribution from the training one. A very similar model was proposed in \cite{schnabel2016recommendations}.

\section{Proposed Approach}
\label{sec:proposedapproach}

As discussed in Section \ref{sec:intro}, we are interested in building a good predictor for recommendation outcomes under random exposure for all the user-product pairs, which we denote as $\hat{y}_{ij}^{rand}$. In our learning setup, we assume that we have access to a large sample $S_c$ from the logging policy $\pi_c$ and a small sample $S_t$ from the randomized treatment policy $\pi_t^{rand}$. To this end, we propose a multi-task objective that jointly factorizes the matrix of observations $y^c_{ij} \in S_c$ and the matrix of observations $y^t_{ij} \in S_t$.

\subsection{Multi-Task Objective For Exposure Policies That Vary Treatments For A Given Set Of Fixed Users}
For simplicity we assume that the same set of users are exposed differently in control/treatment and the exposure change between treatment and control is explained by the difference in product representations in the two policies. In this case, \emph{the users have a fixed representation (either as a set of user features or as a user embedding vector)}. This is the evaluation setup proposed in \cite{liangcausal} and the equivalent evaluation of the work in \cite{schnabel2016recommendations}, where the test sample is produced according to $\pi^{rand}_t$.

The case of predicting outcomes under changing treatment exposures for the same set of users is also addressed by the authors in \cite{rosenfeld2016predicting}. They assume a user-independent exposure policy and derive a multi-task objective that explains jointly the small sample $S_t$ and the big sample $S_c$ and regularizes the difference in the two resulting product representations. In order to motivate this objective, the authors assume that both the expected factual control and treatment rewards can be approximated as linear predictors over the fixed user representations $u_i$:
\begin{equation}
\begin{split}
y^{c}_{ij} \approx <\theta^{c}_j,u_i>\\
y^{t}_{ij} \approx <\theta^{t}_j,u_i>
\end{split}
\end{equation}
where $\theta^{c}_j,\theta^{t}_j$ are the control/treatment vectorial representations of product $j$ and $< x , y >$ denotes the inner product between the vectors $x$ and $y$.

As a result and using the logic of Equation \ref{eq:itepair}, we can approximate the ITE of a user-product pair $i,j$ as the difference between the two:
\begin{equation}
\widehat{ITE}_{ij} = <\theta^t_j,u_i> - <\theta^c_j,u_i> = <\theta^{\Delta}_j, u_i>
\end{equation}

We define the first part of our joint prediction objective as the supervised predictor for $y^t_{ij}$, trained on the limited sample $S_t$. We denote the associated loss term by $l^t_{ij}$:
\begin{equation}
l^t_{ij} =  L(<\theta^t_j, u_i>, y^t_{ij}) + \Omega(\theta^t_j)
\end{equation}

where: $L$ is an arbitrary loss function and and $\Omega(.)$ is a regularization term over the weights of the model.
Switching to matrix notation, we define the associated objective by $L_t$:
\begin{equation}
\label{eq:lt}
L_{t} = \sum_{(i,j,y_{ij}) \in S_t} l^t_{ij} = L(U \Theta_t  , Y_t) + \Omega(\Theta_t)
\end{equation}

where: $\Theta_t$ is the parameter matrix of treatment product representations, $U$ is the fixed matrix of the user representations, $Y_t$ is the observed rewards matrix, $L(.)$ is an arbitrary loss function and $\Omega(.)$ is a regularization term over the weights of the model.

Then, if we want to leverage the ample control data, we can use our treatment product representations by subtracting $ \theta^{\Delta}_j$ since $<\theta^c_j,u_i> = <\theta^t_j,u_i>  - <\theta^{\Delta}_j, u_i>$. We denote the associated objective by $l^{c}_{ij}$:
\begin{equation}
l^c_{ij} = L(<\theta^t_j - \theta^{\Delta}_j, u_i>, y^{c}_{ij}) + \Omega(\theta^t_j) + \Omega(\theta^{\Delta}_j) 
\end{equation}
If instead of parametrizing it in terms of treatment embeddings $\theta^t_j$ and a set of difference vectors $\theta^{\Delta}_j$, we choose to express the loss in terms of the new control embedding $\theta^c_j$, and switching to matrix notation, we have that:

\begin{equation}
\label{eq:lc}
\begin{split}
L_{c} = L(U \Theta_c  , Y_c) + \Omega(\Theta_c) + \Omega(\Theta_t - \Theta_c)
\end{split}
\end{equation}

By putting the two tasks together (eq. \ref{eq:lt} and \ref{eq:lc}), we have the equation of the joint tasks loss $L^{prod}_{CausE}$:
\begin{equation}
\label{eq:prod}
\begin{split}
L^{prod}_{CausE}= \underbrace{L(U \Theta_t  , Y_t) + \Omega(\Theta_t)}_{treatment~task~loss} +  \underbrace{L(U \Theta_c  , Y_c) + \Omega(\Theta_c)}_{control~task~loss} + \\ + \underbrace{\Omega(\Theta_t - \Theta_c)}_{regularizer~between~tasks}
\end{split}
\end{equation}
We can similarly define $L^{user}_{CausE}$, where the user embeddings are learnt over fixed product representations. This type of modelization is relevant for the cases where we know that the treatment policy differs from the control policy in terms of the choice of users.

\subsubsection{Multi-task Objective For Exposure Policies that Vary Both Treatments And Users}
In this case, the change in exposure between treatment and control is explained both by the different choice in exposed users and products shown. To this end, we modify our objective function to allow for the user representations to change, and using similar reasoning to eq (\ref{eq:prod})  we reach $L_{CausE}$, the final loss function for our method:
\begin{equation}
\label{eq:user}
\begin{split}
L_{CausE}= \underbrace{L(\Gamma_t\Theta_t  , Y_t) + \Omega(\Gamma_t,\Theta_t)}_{treatment~task~loss} +  \underbrace{L(\Gamma_c\Theta_c  , Y_c) + \Omega(\Gamma_c,\Theta_c)}_{control~task~loss} + \\ + \underbrace{\Omega(\Gamma_t - \Gamma_c) + \Omega(\Theta_t - \Theta_c) }_{regularizers~between~tasks}
\end{split}
\end{equation}\\
We can recover the previous loss as a special case by setting the regularization parameter on the difference between treatment and control representations for users to infinity.

\subsubsection{Optimization}
We optimize $L_{CausE}$ using Stochastic Gradient Descent (SGD) with momentum \cite{sutskever2013importance} and a linearly decaying learning rate. For our experiments we optimize $L^{prod}_{CausE}$ since our treatment policy depends only on products and not on users. Algorithm \ref{algo:cause} details the general \emph{CausE} method.

\begin{algorithm}[h]

\SetKwInOut{Input}{Input}
\SetKwInOut{Output}{Output}

 \Input{Mini-batches of $S_c = { \{ (u^c_i, p^c_j, \delta^c_{ij}) \} }^{M_c}_{i=1}$ and $S_t = { \{ (u^t_i, p^t_j, \delta^t_{ij}) \} }^{M_t}_{i=1}$ , regularization parameters $\lambda_{t},\lambda_{c}$ for the two joint tasks $L_t$ and $L_c$ and $\lambda_{dist}$ the regularization parameter for the discrepancy between the two representations for products and users, learning rate $\eta$}

 \Output{$\Gamma_t, \Gamma_c, \Theta_t, \Theta_c$ - User and Product Control and Treatment Matrices} 
 Random initialization of $\Gamma_t, \Gamma_c, \Theta_t, \Theta_c$ \;
 \While{not converged}{
  Read batch of training samples\;
      \For{\textbf{each} training sample $s$ in the batch:}{
	  		\If{$s \in S_c$}{	
	  			Lookup the product index $j$ and user index $i$ in $\Theta_c, \Gamma_c$ and \\
	  			Update control product vector: $\theta^c_j \leftarrow \theta^c_j - \eta \nabla L^{prod}_{CausE}$ \\ 
	  			Update control user vector: $\gamma^c_i \leftarrow \gamma^c_i - \eta \nabla L^{user}_{CausE}$ 
	  		}    
	  		\If{$s \in S_t$}{	 
	  			Lookup the product index $j$ and user index $i$ in $\Theta_t, \Gamma_t$ and \\
	  			Update treatment product vector: $\theta^c_j \leftarrow \theta^c_j - \eta \nabla L^{prod}_{CausE}$ \\
	  			Update treatment user vector: $\gamma^c_i \leftarrow \gamma^c_i - \eta \nabla L^{user}_{CausE}$ 
	  		}        
	  	}
   }
   \Return{$\Gamma_t, \Gamma_c, \Theta_t, \Theta_c$}
 \caption{CausE Algorithm: Causal Embeddings For Recommendations}
 \label{algo:cause}
\end{algorithm}

\section{Experiments}
\label{sec:experiments}
The experimental section is organized as follows. Firstly, we describe the evaluation setup, namely, the evaluation task, success metrics, the baselines. and the dataset preprocessing protocol. In the second part, we report results of our experiments on both the MovieLens10M and Netflix datasets.

\subsection{Setup}

\subsubsection{Task: Estimating Treatment Rewards $\mathbf{y^t_{ij}}$}~\\
We compare all of the methods on the task of predicting the outcomes $y^t_{ij}$ under treatment policy $\pi_t$, where all of the methods have available at training time a large sample of observed recommendations outcomes from $\pi_c$ and a small sample from $\pi_t$ (to simulate the large control plus small treatment info coming from exploration or an A/B test). Since the resulting setup is basically a classical conversion-rate prediction problem, we will use the associated metrics, namely Mean-Squared Error (MSE) and Negative Log-Likelihood (NLL). For our final numbers we report their lift (eq \ref{eq:lift}) over the performance of the average predictor $AvgCR$ (e.g. the empirical conversion rate on the test dataset), to ensure that the performance of the predictors evaluated on the test distribution is better than a trivial predictor that has access to the test distribution:
\begin{equation}
    \label{eq:lift}
   lift_x^{metric} =  \frac{metric_x - metric_{AvgCR}}{metric_{AvgCR}} 
\end{equation}
where $metric$ is either MSE or NLL and $x$ is the approach being measured.

In addition, as we are concerned with implicit feedback data which is inherently binary, we also report the area under the receiver operating characteristic curve (AUC) to assess the quality of the predictions made by the various approaches.

\subsubsection{Baselines}~\\
To demonstrate the effectiveness of our proposed approaches, we compare with the following baselines:\\

\textbf{Matrix Factorization Baselines: \\}
\textbf{Bayesian Personalized Ranking (BPR)\\}
To compare our approach against a ranking based method, we use Bayesian Personalized Ranking (BPR) for matrix factorization on implicit feedback data \cite{rendle2009bpr}. In this method, we directly learn user and product representations via matrix factorization, optimized using the BPR-OPT criteria and trained using LearnBPR \cite{rendle2009bpr}.\\

\textbf{Supervised-Prod2Vec (SP2V)\\} As a second factorization baseline we will use a Factorization Machine-like method \cite{rendle2010factorization} that approximates $y_{ij}$ as a sigmoid over a linear transform of the inner-product between the user and product representations:
\begin{equation}
  \label{eq:innerpodbias}
\hat{y}_{ij}=\sigma(\alpha <p_j, u_i> + b_i + b_j + b)
\end{equation}
where: $\sigma$ is the sigmoid function, $\alpha$ is a scaling factor for the user-product similarity score, $b$ is the global bias, $b_i$ and $b_j$ are user and product bias terms.
This method can also be seen as an extension of the Prod2Vec algorithm \cite{prod2vec} for factorizing non-symmetric matrices in the presence of supervised feedback (e.g. negatives are not randomly sampled as they are provided within the data). For this reason, we will denote our supervised baseline as Supervised-Prod2Vec (SP2V).\\

\textbf{ Causal Inference Baselines:\\}
\textbf{Weighted-SupervisedP2V (WSP2V)\\} To test the performance of propensity-based causal inference, we employ the SP2V algorithm on propensity-weighted data, which we denote in our experiments as Weighted-SP2V (WSP2V). The method is similar to the Propensity-Scored Matrix Factorization (PMF) from \cite{schnabel2016recommendations} but with cross-entropy reconstruction loss instead of MSE/MAE. An equivalent method was tested in \cite{liangcausal} where the authors argue that their best performing model, ExpoMF trained causally (CAU) with conditional prediction (Cond), is equivalent to PMF.\\

\textbf{BanditNet (BN)\\}
To utilize BanditNet as a baseline, we use SP2V as our target policy $\pi_w$. For the existing policy $\pi_c$, we model the behavior of the recommendation system as a popularity-based solution, described by the marginal probability of each product in the training data. In our experiments, in order to keep the magnitude of the resulting factor $\frac{1}{\pi_c(p_j)}$ within a reasonable magnitude we replace it with the $\frac{\pi^{rand}(p_j)}{\pi_c(p_j)}$ and cap it to maximum 100. Under this setup, BanditNet with $\lambda=0$ has a similar formulation with WSP2V. As in the original paper \cite{joachims2-18deep}, we train BanditNet using SGD with momentum \cite{sutskever2013importance}. \\

\begin{table*}[t!]
  \centering
  \begin{tabular}{@{}l c c c c c c @{}}
  \toprule
  \textbf{Method}      & \multicolumn{3}{c}{\textbf{MovieLens10M (SKEW)}}  & \multicolumn{3}{c}{\textbf{Netflix (SKEW)}}    \\ \midrule
              & \textbf{MSE lift}  & \textbf{NLL lift} & \textbf{AUC}   & \textbf{MSE lift}     & \textbf{NLL lift}   & \textbf{AUC} \\
  
  \textit{BPR-no}  & $-$  & $-$ & $0.693 (\pm 0.001)$ & $-$ & $-$ & $0.665 (\pm0.001)$   \\
  \textit{BPR-blend}  & $-$  & $-$ & $0.711 (\pm 0.001)$ & $-$ & $-$ & $0.671 (\pm0.001)$   \\
  \textit{SP2V-no}  & $+3.94\% (\pm 0.04)$  & $+4.50\% (\pm 0.04)$ & $0.757 (\pm 0.001)$ & $+10.82\% (\pm0.02)$ & $+10.19\% (\pm0.01)$ & $0.752(\pm0.002)$   \\
  \textit{SP2V-blend} & $+4.37\% (\pm 0.04)$  & $+5.01\% (\pm 0.05)$ & $0.768 (\pm 0.001)$ & $+12.82\% (\pm0.02)$ & $+11.54\% (\pm0.02)$ & $0.764(\pm0.003)$   \\
  \textit{SP2V-test}  & $+2.45\% (\pm 0.02)$  & $+3.56\% (\pm 0.02)$ & $0.741 (\pm 0.001)$ & $+05.67\% (\pm0.02)$ & $+06.23\% (\pm0.02)$ & $0.739(\pm0.004)$   \\
  \textit{WSP2V-no}   & $+5.66\% (\pm 0.03)$  & $+7.44\% (\pm 0.03)$ & $0.786 (\pm 0.001)$ & $+13.52\% (\pm0.01)$ & $+13.11\% (\pm0.01)$ & $0.779(\pm0.001)$   \\
  \textit{WSP2V-blend}  & $+6.14\% (\pm 0.03)$  & $+8.05\% (\pm 0.03)$ & $0.792 (\pm 0.001)$ & $+14.72\% (\pm0.02)$ & $+14.23\% (\pm0.02)$ & $0.782 (\pm0.002)$   \\
  \textit{BN-blend}  & $-$  & $-$ & $0.794 (\pm 0.001)$ & $-$ & $-$ & $0.785 (\pm0.001)$   \\
  \hline
  \hline
  \textit{CausE-avg}  & $+12.67\% (\pm 0.09)$  & $+15.15\% (\pm 0.08)$ & $0.804 (\pm0.001)$ & $+15.62\% (\pm0.02)$ & $+15.21\% (\pm0.02)$ & $0.799 (\pm0.002)$   \\
  \textit{CausE-prod-T}  & $+07.46\% (\pm 0.08)$  & $+10.44\% (\pm 0.09)$ & $0.779 (\pm0.001)$ & $+13.97\% (\pm0.02)$ & $+13.52\% (\pm0.02)$ & $0.789 (\pm0.003)$   \\
  \hline
  \textbf{CausE-prod-C}  & $\mathbf{+15.48\% (\pm 0.09)}$  & $\mathbf{+19.12\% (\pm 0.08)}$ & $\mathbf{0.814 (\pm0.001)}$ & $\mathbf{+17.82\% (\pm0.02)}$ & $\mathbf{+17.19\% (\pm0.02)}$ & $\mathbf{0.821 (\pm0.003)}$   \\
  \bottomrule
  \end{tabular}
  \caption{Results for MovieLens10M and Netflix on the Skewed (SKEW) test datasets. All three versions of the \emph{CausE} algorithm outperform both the standard and the IPS-weighted causal factorization methods, with \textit{CausE-avg} and \textit{CausE-prod-C} also out-performing BanditNet. We can observe that our best approach \emph{CausE-prod-C} outperforms the best competing approaches \emph{WSP2V-blend} by a large margin (21\% MSE and 20\% NLL lifts on the MovieLens10M dataset) and \textit{BN-blend} (5\% AUC lift on MovieLens10M).}
  \label{tab:results}
  \end{table*}

\subsubsection{Leveraging The Exploration Sample $\mathbf{S_t}$}~\\
In order to better understand the leverage we obtain by transferring information from the limited treatment policy sample, we define four possible setups of incorporating the exploration data: 
\begin{itemize}
\item No adaptation \textit{(no)} - the algorithm does not take into account the exploration data, it is trained only on the $S_c$ sample.
\item Blended adaptation \textit{(blend)} - the algorithm is trained on the union of the $S_c$ and $S_t$ samples.
\item Test-only adaptation \textit{(test)} - the algorithm is trained only on the $S_t$ samples.
\item Average test adaptation \textit{(avg)} - the algorithm constructs an average treatment product by pooling all of the points from the $S_t$ sample into a single vector (it applies only to \emph{CausE}).
\item Product-level adaptation \textit{(prod)} - the algorithm constructs a separate treatment embedding for each product based on the $S_t$ sample (it applies only to \emph{CausE}). For the final prediction we can use either the control (denoted by \textit{CausE-prod-C}) or the treatment product representation (denoted by \textit{CausE-prod-T}).
\end{itemize}

In the following sub-section we detail our methodology for creating the evaluation datasets.\\

\subsubsection{Datasets: MovieLens10M and Netflix}~\\
The datasets used for evaluation are the MovieLens10M \cite{harper2016movielens} (with 71567 unique users and 10677 unique products) and Netflix \cite{bennett2007netflix} (with 480189 unique users and 17770 unique products) recommendation datasets. We also use the MovieLens100K dataset \cite{harper2016movielens} to explore how changes in the quantity of test set injected into the training set affect performance. In order to be able to simulate causal effects under uniform exposure, all datasets are transformed using the standard protocol introduced by Liang \etal \cite{liangcausal}: First we binarize the ratings $y_{ij}$ by setting all observed five-star ratings to one (click) and everything else to zero (view only). We then create two datasets: regular (REG) and skewed (SKEW), each one with 70/10/20 train/validation/test event splits.

To create the regular dataset, for each user we split the products in train/validation/test, similarly with the normal supervised evaluation setup. Whilst in the creation of the skewed dataset, we first generate the test dataset where we expose as uniformly as possible each user to each product. This simulates a random test recommendation policy $\pi_t^{rand}$, which the authors argue is the best policy to collect unbiased feedback for training a recommendation model. In order to achieve this we estimate the popularity of each product and find the least popular product. For all other products we compute how much more common it is than the least popular product. This can then used to compute an acceptance probability e.g. if a product is 100 times more popular than the least popular product we accept these products in the test set with 1\% probability.

In the original protocol, the skewed training and validation sets partition the rest of the data to obtain the final 70/10/20 split. This setup is sufficient to be able to compute the propensity weights between the training policy and the test policy, e.g. $\frac{\pi_t(p_j|u_i)}{\pi_c(p_j|u_i)}$ that will be used by WP2V. However, for \emph{CausE} we need at training time an explicit sample from $\pi_t$ and not only the propensity ratio. Because of this, we increase the size of the uniform test to 30\%, 20\% used for final testing and 10\% to be injected in the training data.

To recap, the final skewed dataset for which we report the results in Table \ref{tab:results} comprises three parts: 70\% train (60\% from $\pi_c$ and 10\% from $\pi_t$), 10\% validation (all from $\pi_c$), 20\% test (all from $\pi_t$). To limit the number of test products that are not available in training, due to fully sampling them for skewed testing, we cap the maximum probability for a product event to be moved in the test split to 0.9. Additionally, to help limit overfitting for WP2V, we cap the maximum propensity weight to 10.

\subsection{Results}
Table \ref{tab:results} shows the results for all approaches evaluated on the binarized MovieLens10M and Netflix datasets. 

We use the standard cross-entropy loss as the loss function $L$. In terms of implementation, all the baseline and \emph{CausE} approaches tested have been coded in Tensorflow and were optimised using SGD with momentum and a linearly decaying learning rate. 

For our choice of regularization functions, similarly with the previous work of \cite{shalit2016estimating} and \cite{rosenfeld2016predicting}, we experimented with both $L_1$ and $L_2$ for penalizing the embedding matrices and the discrepancy between the two product representations associated with the $S_c$ and $S_t$ distributions. We report our best results which we reached with the configuration $\Omega_{t,c}(.)=L_2$ and $\Omega_{dist}(.)=L_1$.

\begin{figure*}[!t]
  \centering
  \begin{subfigure}[b]{0.32\textwidth}
    \includegraphics[width=\linewidth]{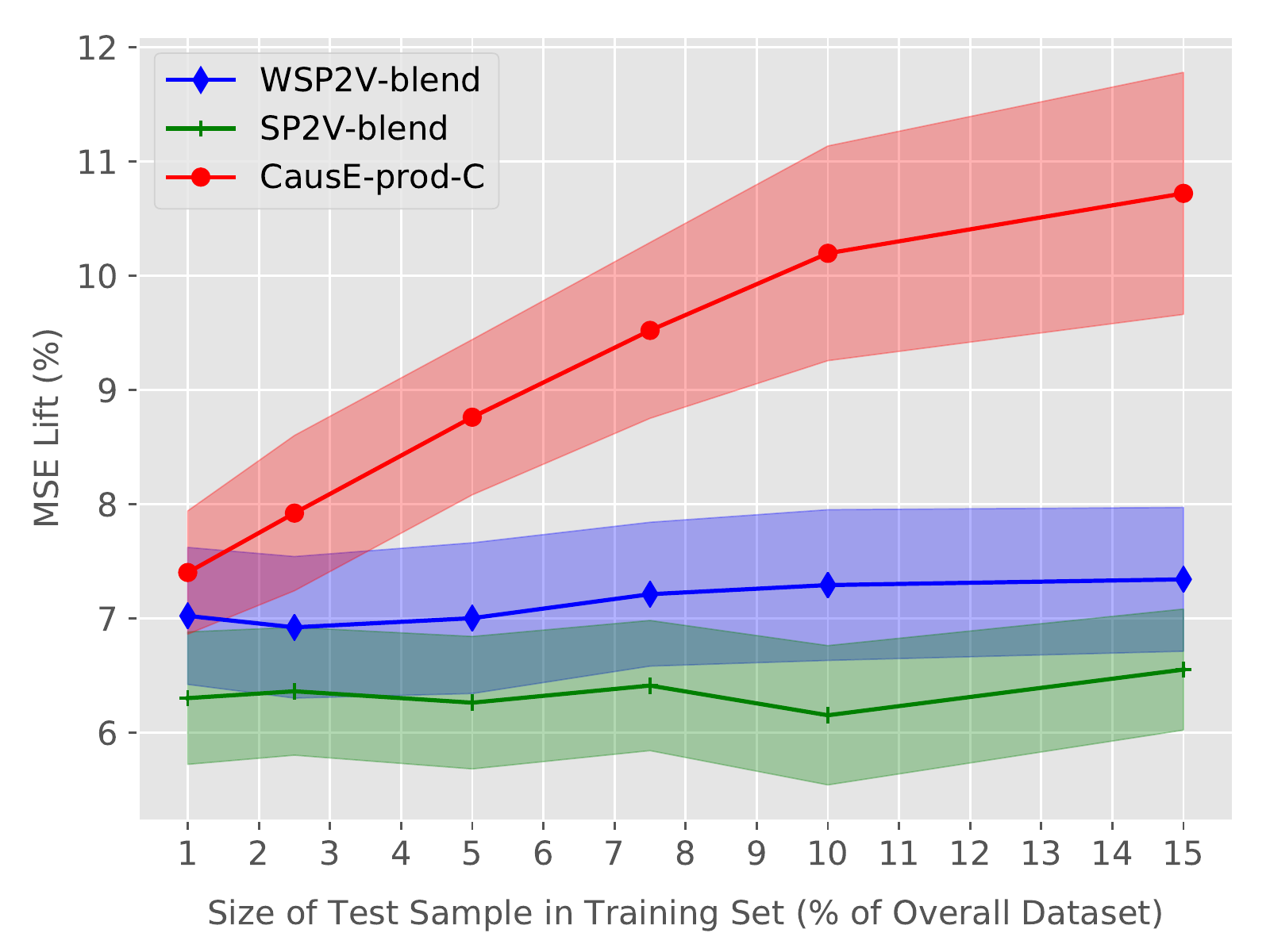}
      \caption{MSE Lift}
      \label{fig:mse}
  \end{subfigure}
  \hfill
  \begin{subfigure}[b]{0.32\textwidth}
    \includegraphics[width=\linewidth]{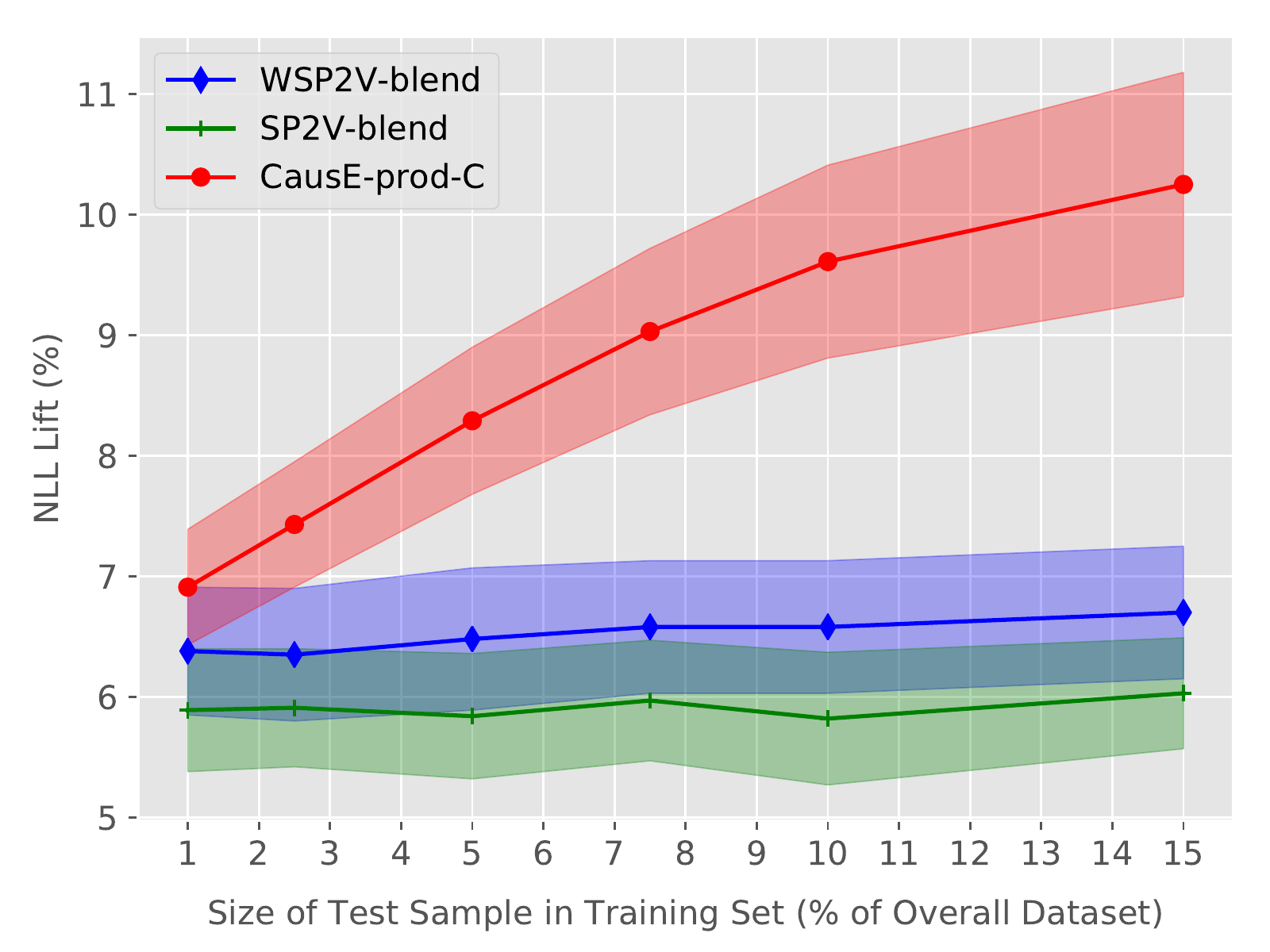}
      \caption{NLL Lift}
      \label{fig:nll}
  \end{subfigure}
  \hfill
  \begin{subfigure}[b]{0.32\textwidth}
    \includegraphics[width=\linewidth]{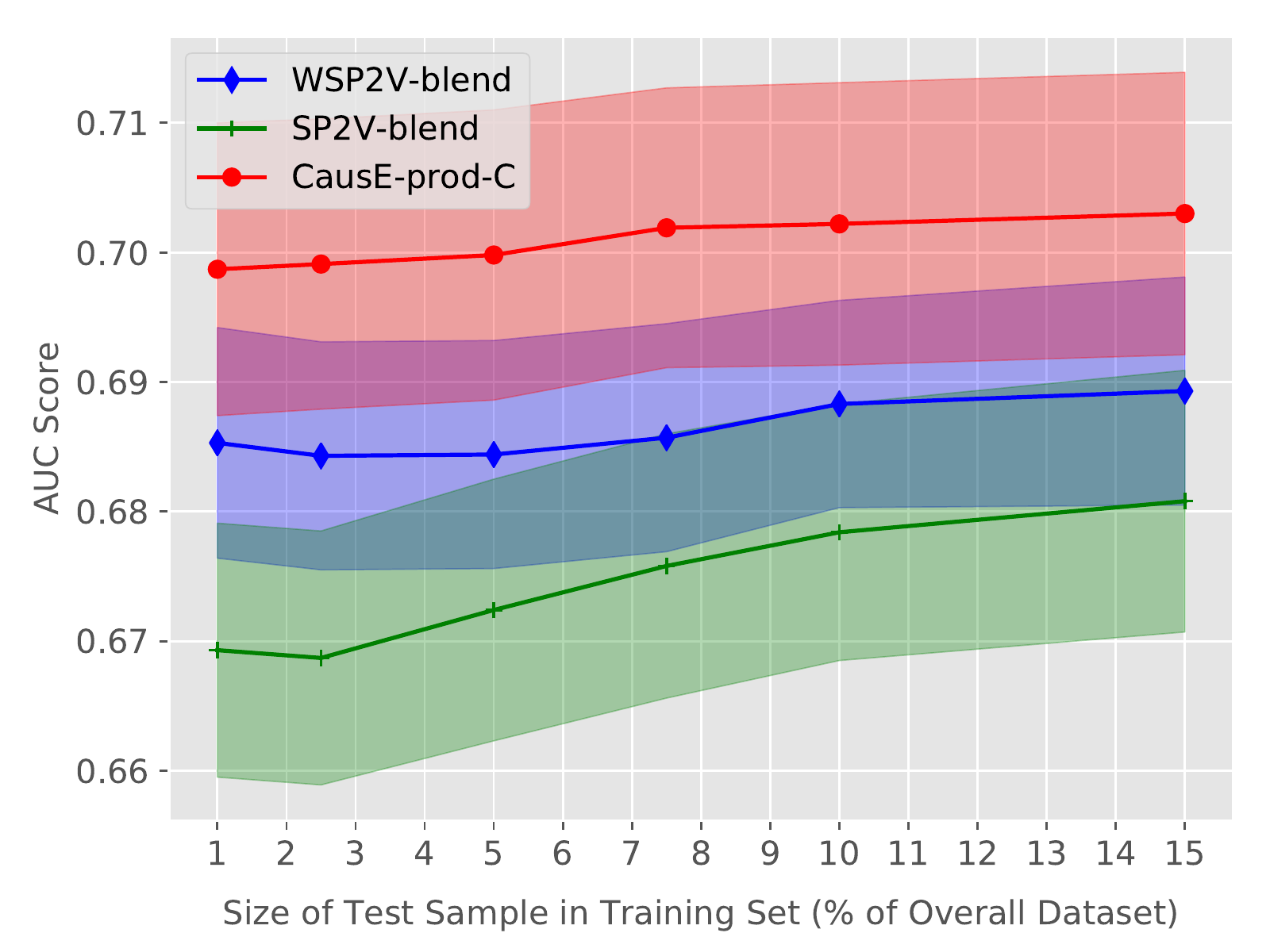}
      \caption{AUC}
      \label{fig:auc}
  \end{subfigure}
  \hfill
  \vspace{-0.3cm}
  \caption{Change in MSE (\ref{fig:mse}), NLL (\ref{fig:nll}) and AUC (\ref{fig:auc}) on MovieLens100K as more test set is injected into the blend training dataset.}\label{fig:control}
\end{figure*}

As one would expect, in the case of the REG dataset, where the training and test distributions are constant, all methods demonstrated similar performance. For the sake of brevity we do not present the REG results in Table \ref{tab:results}, but for the MovieLens10M dataset all approaches achieved an MSE lift of $+8\% (\pm0.02)$.

On the SKEW datasets, where the test distribution is very different from the training distribution, the standard matrix factorization and supervised methods (\emph{SP2V} and \emph{BPR}) perform badly, with the method having access to the $S_t$ sample during training time showing a marginal increase in performance (\emph{SP2V-blend}). \emph{SP2V-test} is the only method to be trained exclusively upon data sampled from the test distribution, but as expected, it cannot perform as well as other methods due to sparsity issues arising from its comparatively smaller size. 

As expected, the competing causal approaches lead to better results: the IPS-weighted methods (\emph{WSP2V-no/blend}) can take advantage of the propensity information and are able to outperform the standard supervised methods and manage to exhibit some domain adaptation from $S_c$ to $S_t$ on both datasets. BanditNet (\emph{BN-blend}) outperforms the IPS-weighted approaches and demonstrates performance on the test distribution that is the closest to our proposed method.  

Finally, our proposed \textit{CausE} method significantly outperforms all baselines across both datasets, demonstrating that it has a better capacity to leverage the small test distribution sample $S_t$. We observe that, out of the three \emph{CausE} variants,  \emph{CausE-prod-C}, the variant  that is using the regularized control matrix, clearly out-performs the others. We believe that this is due to the fact that fitting the larger training sample $S_c$ allows to model correctly the personalized user responses, while controlling the deviation from the target task via the discrepancy regularizer.

In terms of the impact of the size of the $S_t$ sample on performance, figure \ref{fig:control} shows that \emph{CausE} is able to leverage the extra data from the test distribution much better than the other methods, giving it the best overall lift.

\section{Conclusions}
\label{sec:conclusion}

In this paper, we have introduced a novel method for factorizing matrices of user implicit feedback that optimizes for causal recommendation outcomes. We show that the objective of optimizing for causal recommendations is equivalent with factorizing a matrix of user responses collected under uniform exposure to item recommendations and propose \emph{CausE}, a domain adaptation-inspired method that learns from a large sample of biased exposure feedback data $S_c$ and a small sample of unbiased exposure feedback data $S_t$. The resulting method is a simple extension to current matrix factorization algorithms that adds a regularizer term on the discrepancy between the item vectors used to fit the biased sample $S_c$ and their counter-part representations that fit the uniform exposure sample $S_t$.  

Since, unlike other competing causal models, we are able to leverage the large sample of biased feedback data, we show that we can clearly outperform both classical matrix factorization methods, in addition to recent causal approaches such as IPS-weighted factorization methods and BanditNet. 

We believe this to be promising for industrial applications for two main reasons: Firstly, in real-world scenarios, most of the current user feedback data is collected under a biased recommendation system which fits perfectly in our approach. Secondly, most of real-world recommender systems are based on matrix factorization, which would require little modifications in order to be able to experiment with our proposed approach.

As future work, we plan to extend this approach to user sequence modeling and leverage both organic user activity sequences and influenced user activity sequences in order to make sequential product recommendation policies more incremental.

\section*{Acknowledgements}

The authors would like to thank David Rohde, Mike Gartrell and Jeremie Mary for their insightful feedback and comments during the preparation of this manuscript.   

\bibliographystyle{ACM-Reference-Format}
\bibliography{literature}

\end{document}